\documentclass{article}
\pdfoutput=1
\usepackage[a4paper, total={6in, 9in}]{geometry}
\usepackage{hyperref}
\usepackage[T1]{fontenc}
\usepackage{lmodern}

\usepackage{url}
\usepackage{booktabs}
\usepackage{framed,multirow}
\usepackage{microtype}
\usepackage{amsmath}
\usepackage{amssymb}
\usepackage{amsfonts}
\usepackage{amsthm}
\usepackage{amsopn}
\usepackage{bm} 
\usepackage{graphicx}
\usepackage{amsfonts}
\usepackage{longtable}
\usepackage{epsf, epsfig}
\usepackage{epstopdf}
\usepackage{subfigure}
\usepackage{caption}
\usepackage{algorithm}
\usepackage[table]{xcolor}
\usepackage{multirow}
\usepackage{diagbox}
\usepackage{footnote}
\usepackage{bbding}
\usepackage{xargs} 
\usepackage{xspace}
\usepackage{diagbox}
\usepackage{xr}
\usepackage{authblk}
\usepackage{paralist}
\usepackage{titling}
\usepackage{nicefrac}
\usepackage{crimson}
\usepackage{caption}
\usepackage{algorithmic,algorithm}

\newcommand{\vct}[1]{\boldsymbol{#1}} 
\newcommand{\mat}[1]{\boldsymbol{#1}}

\DeclareMathOperator{\argmin}{arg\,min}

\newcommand{\draft}[1]{}
\pretitle{\begin{center} \LARGE \bf}
\posttitle{\par\end{center}\vskip 0.5em}
\preauthor{
    \begin{center}
    \normalsize \lineskip 0.5em%
    \begin{tabular}[t]{c}
}
\postauthor{\end{tabular}\par\end{center}}
\predate{\begin{center}\large}
\postdate{\par\end{center}}

\newcommand{\cross}[1]{#1 $\times$ #1}

\title{Simultaneous reconstruction of the initial pressure and sound speed in photoacoustic tomography using a deep-learning approach}
\author{Hongming Shan, Christopher Wiedeman, Ge Wang, Yang Yang}
\affil{
    Contact: \texttt{\normalsize shanh@rpi.edu, wiedec@rpi.edu, wangg6@rpi.edu, yangy5@msu.edu}
}
\affil{Rensselaer Polytechnic Institute\\Michigan State University}

\date{\today}

\begin{document}

\maketitle

\begin{abstract}
Photoacoustic tomography seeks to reconstruct an acoustic initial pressure distribution from the measurement of the ultrasound waveforms. Conventional methods assume a-prior knowledge of the sound speed distribution, which practically is unknown.  One way to circumvent the issue is to simultaneously reconstruct both the  acoustic initial pressure and speed. In this article, we develop a novel data-driven method that integrates an advanced deep neural network through model-based iteration.  The image of the initial pressure is significantly improved in our numerical simulation. 
\end{abstract}

\section{Introduction}

Photoacoustic tomography (PAT) is an emerging non-invasive modality that has manifested an enormous prospect for clinical practices \cite{xia2014photoacoustic}. In PAT, the tissue is illuminated with near-infrared light of wavelength 650-900nm. The absorbed optical energy is transformed into acoustic energy through the photoacoustic effect, and the generated ultrasound is measured by transducer arrays around the tissue in order to retrieve the optical internal properties. The coupling mechanism of the optical and ultrasound waves gives multiple advantages over conventional standalone modalities. For instance, the acoustic wave experiences less scattering in tissue compared to optical wave, allowing PAT to break the optical diffusion limit and generate high-resolution images~\cite{Wang2010breaking} while preserving intrinsic optical contrast~\cite{Yao2011photoacoustic}.

Typical photo-acoustic signal generation comprises three steps: (1) the tissue absorbs light; (2) the absorbed optical energy heats the tissue and raises the temperature; (3) thermo-elastic expansion occurs and generates ultrasound. The image formation in PAT is to recover the distribution of the deposited energy, known as the local optical fluence, from the ultrasound signals that are recorded by the sensors deployed around the tissue. As the initial ultrasound pressure is approximately proportional to the optical fluence, it is sufficient to reconstruct the initial pressure from the recorded ultrasound signals. Conventional reconstructive schemes, such as back-projection based methods~\cite{finch2009recovering, finch2007inversion, haltmeier2005filtered, kruger1999thermoacoustic, kruger2003thermoacoustic, kunyansky2007explicit, xu2005universal, xu2004reconstructions} and time-reversal based methods~\cite{chung2014neumann, palacios2016reconstruction, homan2013multi, hristova2009time,  hristova2008reconstruction, qian2011efficient, stefanov2009thermoacoustic, stefanov2011thermoacoustic, stefanov2015multiwave, stefanov2017thermo, stefanov2016multiwave, belhachmi2016direct, javaherian2019direct}, typically assume the sound speed is precisely known. This assumption however is not always fulfilled, and the precise sound speed distribution is often unknown. It is therefore of practical significance to study simultaneous reconstruction of both the sound speed distribution and the acoustic initial pressure.

The purpose of this article is to address this issue using a data-driven approach. Data-driven approaches such as machine learning and deep learning have gained increasing attention recently in medical image reconstruction~\cite{wang2016perspective,wang2018image}. The powerful computational capacity of modern super computers has enabled these methods to enhance medical image reconstruction by extracting the hidden patterns in the data that would otherwise be lost. In PAT, data driven methods have achieved state-of-the-art results in PAT image reconstruction in the scenarios of sparse data ~\cite{Antholzer2019sparse, Antholzer2018DL, Antholzer2018L1}, limited view~\cite{Hauptmann2018model, Waibel2018reconstruction, schwab2019deep}, artifacts removal~\cite{guan2019fully, Allman2018photoacoustic, allman2018exploring}, as well as other applications~\cite{kelly2017deep,Schwab2018realtimel,shan2019accelerated}.
In addition to PAT, data-driven approaches have also found broad application in the image reconstruction of many other imaging modalities such as Computed Tomography (CT), Magnetic Resonance Imaging (MRI), as well as other related fields like radiomics and radiotheraphy, see~\cite{chen2018learn,shan2019competitive,shan20183d,you2018ct,xie2019dual,fan2019quadratic,shan2019novel, lyu2019mri, shan2017enhancing,lei2018soft, peng2019mcdnet,huynh2016digital}
for some examples and applications along these directions.

\section{Preliminary}

In this section, we formulate the mathematical model of the simultaneous reconstruction problem in PAT and review the recent advances based on model-driven approaches.

Propagation of ultrasound in PAT is typically modeled as an initial value problem for the wave equation. This can be written as
\begin{equation} \label{eq:forward}
\left\{
\begin{array}{rcl}

\rho(\vct{r})\frac{\partial \vct{u}}{\partial t}(\vct{r}, t) + \nabla p(\vct{r}, t) & = & \vct{0}, \quad\quad\quad \text{ in } (0,T)\times\mathbb{R}^d\\

\frac{1}{\rho(\vct{r})c(\vct{r})^2} \frac{\partial p}{\partial t}(\vct{r}, t) + \nabla \cdot \vct{u}(\vct{r}, t) & = & 0, \quad\quad\quad \text{ in } (0,T)\times\mathbb{R}^d\\

p(\vct{r}, 0) & = & p_0(\vct{r}), \\

\vct{u}(\vct{r}, 0) & = & \vct{0},

\end{array}
\right.
\end{equation}
where $d=2,3$ is the dimension, $\vct{u}$ is the acoustic velocity, $p$ is the acoustic pressure, $T>0$ is the stoppage time of the measurement, $\rho$ is the mass density, $c$ is the sound speed, and $p_0$ is the initial pressure distribution. We remark that another popular model in PAT is to deal with the second order acoustic equation rather than the above first order system. These models are more or less equivalent under suitable assumptions.

Let $\Omega$ be a bounded domain in $\mathbb{R}^d$ representing the tissue, with the boundary of $\Omega$ denoted by $\partial\Omega$. The measurement in PAT is the detected ultrasound signals, that is the temporal boundary data $p|_{[0,T]\times\partial\Omega}$. We further introduce a \textit{measurement map} $\Lambda$ which is defined as
\begin{equation} \label{eq:Lambda}
\Lambda: (c,p_0) \mapsto p|_{[0,T]\times\partial\Omega}
\end{equation}
where $p$ is the solution of~\eqref{eq:forward}. Assuming the mass density $\rho$ is known, then the simultaneous reconstruction problem in PAT can be recast as the inverse problem to recover the pair $(c,p_0)$ from $p|_{[0,T]\times\partial\Omega}$; in other words, to invert the operator $\Lambda$. Note that $\Lambda$ is linear in $p_0$ but non-linear in $c$.

We recall some known results regarding the simultaneous reconstruction in PAT. Linearization of the map~\eqref{eq:Lambda} at sufficiently smooth pairs is studied in~\cite{stefanov2013instability} and the linearized inversion is shown to be unstable in any scale of Sobolev spaces. Invertibility of $\Lambda$ is established in~\cite{finch2013transmission} for radially symmetric sound speeds in any odd dimension higher than three; in~\cite{liu2015jointdetermination, knox2018jointdetermination} for sound speeds that satisfy certain orthogonality relation with harmonic functions. Numerical simulations have been implemented in~\cite{yuan2006simultaneous} using the finite element model, and in~\cite{matthews2018parameterized, zhang2008simultaneous, matthews2017joint} for parametrized sound speeds.

Our data-driven approach is inspired by the work of Matthews et al.~\cite{matthews2018parameterized}, where the authors proposed a direct optimization approach towards the simultaneous reconstruction, with the sound speed being confined in a parametrized lower-dimensional space. We denote the measurement by $g$ for simplicity. With slight abuse of notations, the discretized versions of $\vct{p}_0$ and $\vct{c}$ are still denoted by themselves, which are two matrices. For a fixed sound speed $c$, the operator $\Lambda(c,\cdot)$ is linear, hence its discretization is another matrix, say $H(c)$. The idea in Ref~\cite{matthews2018parameterized} is to minimize the objective function
\begin{equation}\label{eq:objective_function}
\widehat{\vct{p}_0}, \widehat{\vct{c}} = \argmin_{\vct{p}_0\geq 0, \vct{c} \geq 0} F(\vct{p}_0, \vct{c}) + \beta R(\vct{p}_0)
\end{equation}
where the data fidelity term is defined as $F(\vct{p}_0, \vct{c}) := \frac{1}{2}\| g - \mat{H(c)}\vct{p}_0\|^2$ and $R$ denotes the regularization term for the initial pressure. This objective function can be solved using the proximal optimization method that allows constraints and nonsmooth regularization function for the initial pressure distribution. The initial pressure distribution and sound speed distribution are updated iteratively using gradient descent. The algorithm in Ref~\cite{matthews2018parameterized} is summarized in Algorithm~\ref{alg:conventional} below.
\begin{algorithm}
\caption{Simultaneous Reconstruction by iterative algorithm~\cite{matthews2018parameterized}}\label{alg:conventional}
\begin{algorithmic}[1]
\REQUIRE ${p}_0^{(0)}$, $\vct{c}^{(0)}$, $\beta$
\ENSURE $\widehat{{p}_0}$, $\widehat{\vct{c}}$
\STATE $k \gets 0$
\STATE \textbf{while} stopping criterion is not satisfied \textbf{do}
\STATE \quad Calculate gradients $\nabla_{\vct{p}_0}F$ and $\nabla_{\vct{c}}F$
\STATE \quad Choose step length $\alpha_k^p$ and $\alpha_k^c$
\STATE \quad ${p}_0^{(k+1)} = \mathrm{prox}_{\alpha_k^p \beta R}\big( {p}_0^{(k)} - \alpha_k^p  \nabla_{\vct{p}_0}F \big)$
\STATE \quad ${c}^{(k+1)} =  {c}^{(k)} - \alpha_k^c  \nabla_{\vct{c}}F $
\STATE \quad $k \gets k + 1$
\STATE \textbf{end while}
\STATE  $\widehat{{p}_0} \gets {p}_0^{(k+1)} $
\STATE $\widehat{\vct{c}} \gets {c}^{(k+1)}$ 
\end{algorithmic}
\end{algorithm}

\section{Simultaneous Reconstruction via Deep Learning}

We describe our data-driven method in this section. The motivation stems from some limitations of the iterative algorithm~\ref{alg:conventional}. For instance, repeated selection of step sizes to update the initial pressure and sound speed is time-consuming; selection of the parameter $\beta$ and the regularization term $R$ is highly empirical; tuning $\beta$ can be tedious. To overcome these limitations, we propose a simultaneous reconstruction network (SR-Net) to update the initial pressure and the sound speed for each iteration. 

\begin{figure}[htbp]
    \centering
    \includegraphics[width=1.0\linewidth]{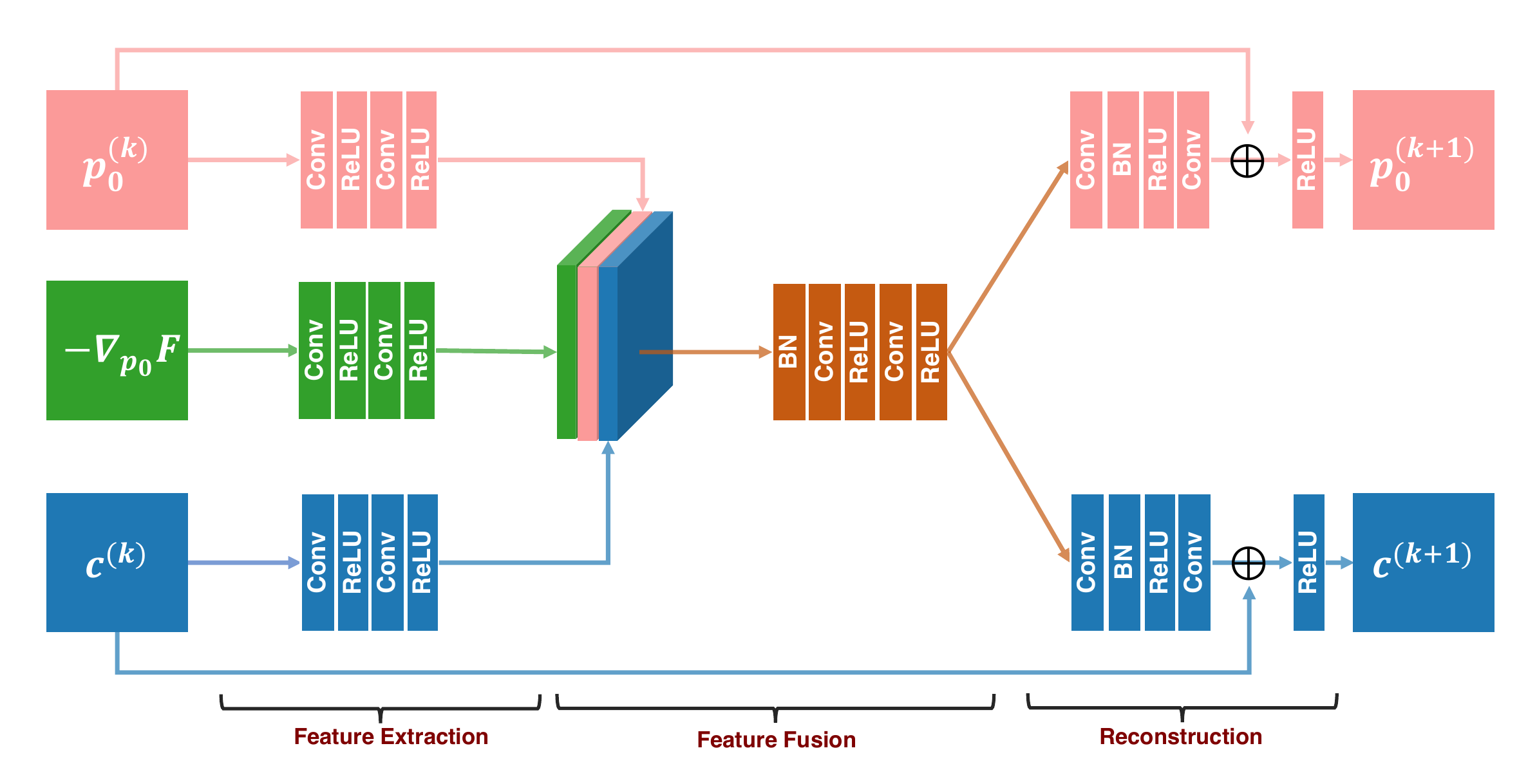}
    \caption{The proposed simultaneous reconstruction network (SR-Net) for PAT. }
    \label{fig:workflow}
\end{figure}

The proposed SR-Net is shown in Fig. \ref{fig:workflow}. It contains three steps - feature extraction, feature fusion, and reconstruction. 
\begin{enumerate}
	\item \textbf{Feature extraction:} We used two convolutional layers to extract features from the initial pressure distribution, the negative gradient of data fidelity with respect to initial pressure, and current sound speed distribution.  
	\item \textbf{Feature fusion:} The feature maps extracted from above three branches are concatenated along the channel dimension, which are then fed into the batch normalization (BN) layer. Since the feature maps from the initial pressure and the sound speed are in different scale, this batch normalization layer can normalize feature maps into the same scale for subsequent feature fusion. The feature fusion part also contains two convolutional layers followed by ReLU. 
	\item \textbf{Reconstruction:} After the feature fusion, we use two convolutional layers to reconstruct the updated initial pressure and sound speed, respectively. A batch normalization layer is used after first convolutional layer to scale the fused feature back to the original scale. A skip connection after the second convolutional layer enables the network to learn the residual just like the traditional iterative algorithm (see step 5-6 in Algorithm~\ref{alg:conventional}). The last ReLU activation function guarantees the updated initial pressure and speed distribution are non-negative. 
\end{enumerate}

\begin{table}[htbp]
	\centering
	\caption{Network architectures of feature extraction, feature fusion, and reconstruction in SR-Net. }\label{network}
	\begin{tabular}{cccc}
	\toprule 
	\textbf{Layer} 	& \textbf{Feature Extraction} $\times3$
	& \textbf{Feature Fusion} & \textbf{Reconstruction} $\times2$\\ \midrule
	1	& \cross{3} {} {} {} {} \textbf{conv}, 32, stride 1 	& \textbf{Batch Normalization}	& \cross{3} {} {} {} {} \textbf{conv}, 16, stride 1 \\ 
	2	&\cross{3} {} {} {} {} \textbf{conv}, 32, stride 1 	& \cross{3} {} {} {} {} \textbf{conv},  64, stride 1	& \textbf{Batch Normalization} \\ 
	3	& & \cross{3} {} {} {} {} \textbf{conv}, 32, stride 1	& \cross{3} {} {} {} {} \textbf{conv}, {} 1, stride 1	\\ 
	
	\bottomrule
	\end{tabular}
\end{table}

A detailed network structure is shown in Table \ref{network}. Note that zero-padding is used  such that all feature maps have the same as the inputs. For each iteration, we trained SR-Net using the simulated dataset. At the $k$-th iteration,  the loss function is chosen as follows:
\begin{equation}
    \begin{aligned}
	\min_{\vct{\theta}_{\mathrm{SR}}} \quad & \mathbb{E}_{(p_0^{(k)}, -\nabla_{\vct{p}_0}F, c^{(k)}, p_0, c)}\bigg[ \big| \vct{p}_0^{(k+1)} - \vct{p}_0 \big| + \frac{1}{1000} \big| \vct{c}^{(k+1)} - \vct{c}\big| \bigg] \\
	\textrm{s.t.}\quad & {p}_0^{(k+1)}, {c}^{(k+1)} = \mathrm{\text{SR-Net}}({p}_0^{(k)},- \nabla_{\vct{p}_0}F, {c}^{(k)})
	\end{aligned}
\end{equation}
where $\theta_{\mathrm{SR}}$ represents the parameters of the SR-Net. $p_0$ and $c$ are the ground-truth labels. We employ the Adam algorithm~\cite{kingma2014adam} to update the parameters in $\theta_\mathrm{SR}$. The gradients of the parameters are computed using a back-propagation algorithm. The iterative reconstruction algorithm is summarized in Algorithm \ref{alg:dl}.

\begin{algorithm}
\caption{Simultaneous reconstruction via Deep learning}\label{alg:dl}
\begin{algorithmic}[1]
\REQUIRE ${p}_0^{(0)}$, $\vct{c}^{(0)}$, $k_{max}$
\ENSURE $\widehat{{p}_0}$, $\widehat{\vct{c}}$
\STATE $k \gets 0$
\STATE \textbf{while} $k < k_{max}$ \textbf{do}
\STATE \quad Calculate gradient $\nabla_{\vct{p}_0}F$
\STATE \quad ${p}_0^{(k+1)}, {c}^{(k+1)} = \mathrm{\text{SR-Net}}({p}_0^{(k)},- \nabla_{\vct{p}_0}F, {c}^{(k)})$
\STATE \quad $k \gets k + 1$
\STATE \textbf{end while}
\STATE  $\widehat{{p}_0} \gets {p}_0^{(k+1)} $
\STATE $\widehat{\vct{c}} \gets {c}^{(k+1)}$ 
\end{algorithmic}
\end{algorithm}

Note that our proposed network is different from the model in Ref~\cite{Hauptmann2018model} because 1) our network aims to reconstruct the initial pressure distribution and sound speed distribution simultaneously, while the model in Ref~\cite{Hauptmann2018model} assumes the sound speed is precisely known; 2) our network is a 2D network, while the model in Ref~\cite{Hauptmann2018model} is a 3D network; and 3) the proposed network used the batch normalization layer to scale the feature maps, while the model in Ref~\cite{Hauptmann2018model} used an explicit coefficient to represent the step length. 

\begin{figure}[h]
    \centering
    \includegraphics[width=1.0\textwidth]{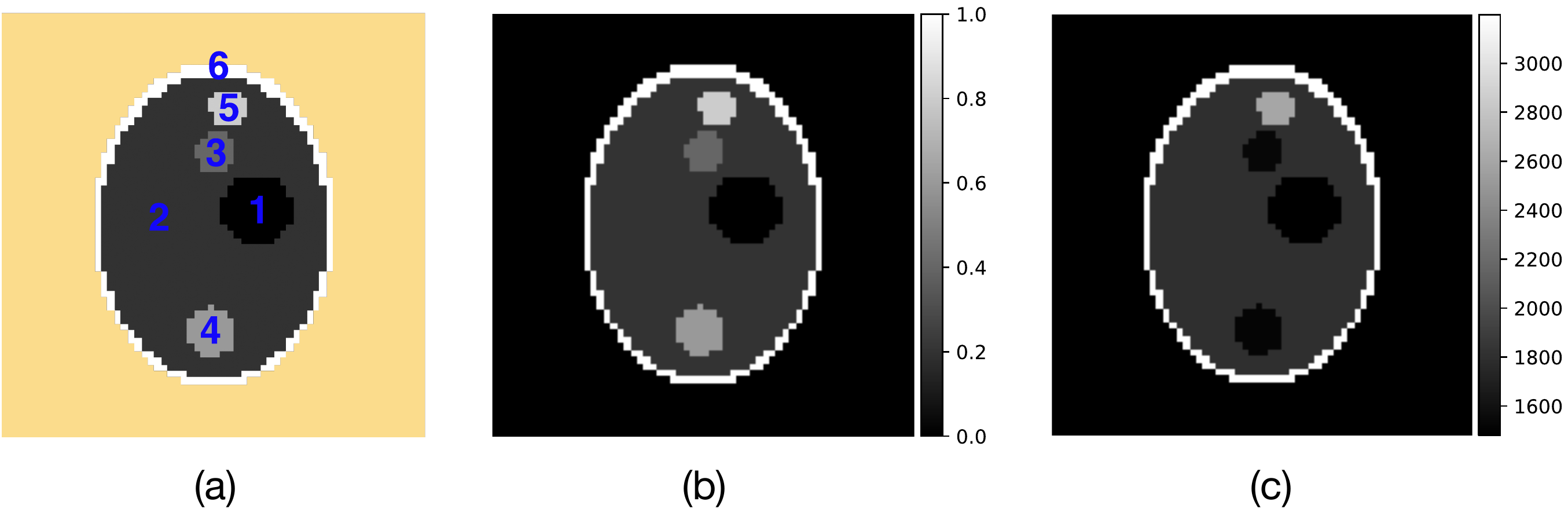}
    \caption{An example of phantom, initial pressure and sound speed. The phantom used in this study contains 6 regions (a), which was filled with different initial pressure shown in (b). The corresponding sound speed distribution is shown in (c). Note that the yellow background in (a) denotes the outside of the phantom. }
    \label{fig:phantom}
\end{figure}

\section{RESULTS}
\subsection{Data generation}

In this study, we used a simplified Shepp-logan phantom to train and test our network. Fig. \ref{fig:phantom} shows the phantom along with the initial pressure and the sound speed distribution. We randomly changed the sizes and positions of the region 1, 3, 4, and 5 in order to increase the diversity of dataset. The corresponding initial pressure and sound speed are shown in Table \ref{tab:index}. We put the phantom into a $64\times 64$ image, which is surrounded by 252 detectors in the rectangle form. For the training purpose, we randomly generated 5,120 images with acoustic signal of the size $252\times 652$. The mass density is assumed to be constant 1 throughout. Then we selected 1,024 images randomly to test the trained model.

\begin{table}[h]
    \centering
    \caption{The initial pressure and sound speed for each region.}
    \begin{tabular}{ccc}
    \toprule 
    \textbf{Index} & \textbf{Initial Pressure} & \textbf{Sound Speed} [m/s] \\ 
    \midrule
    1   &  0.0 & 1480 \\
    2   &  0.2 & 1800 \\
    3   & 0.4 & 1530 \\
    4   & 0.6 & 1520 \\
    5   & 0.8 & 2600 \\
    6   & 1.0 & 3198 \\
    \bottomrule
    \end{tabular}
    
    \label{tab:index}
\end{table}

\begin{figure}[htbp]
    \centering
    \includegraphics[width=1.0\textwidth]{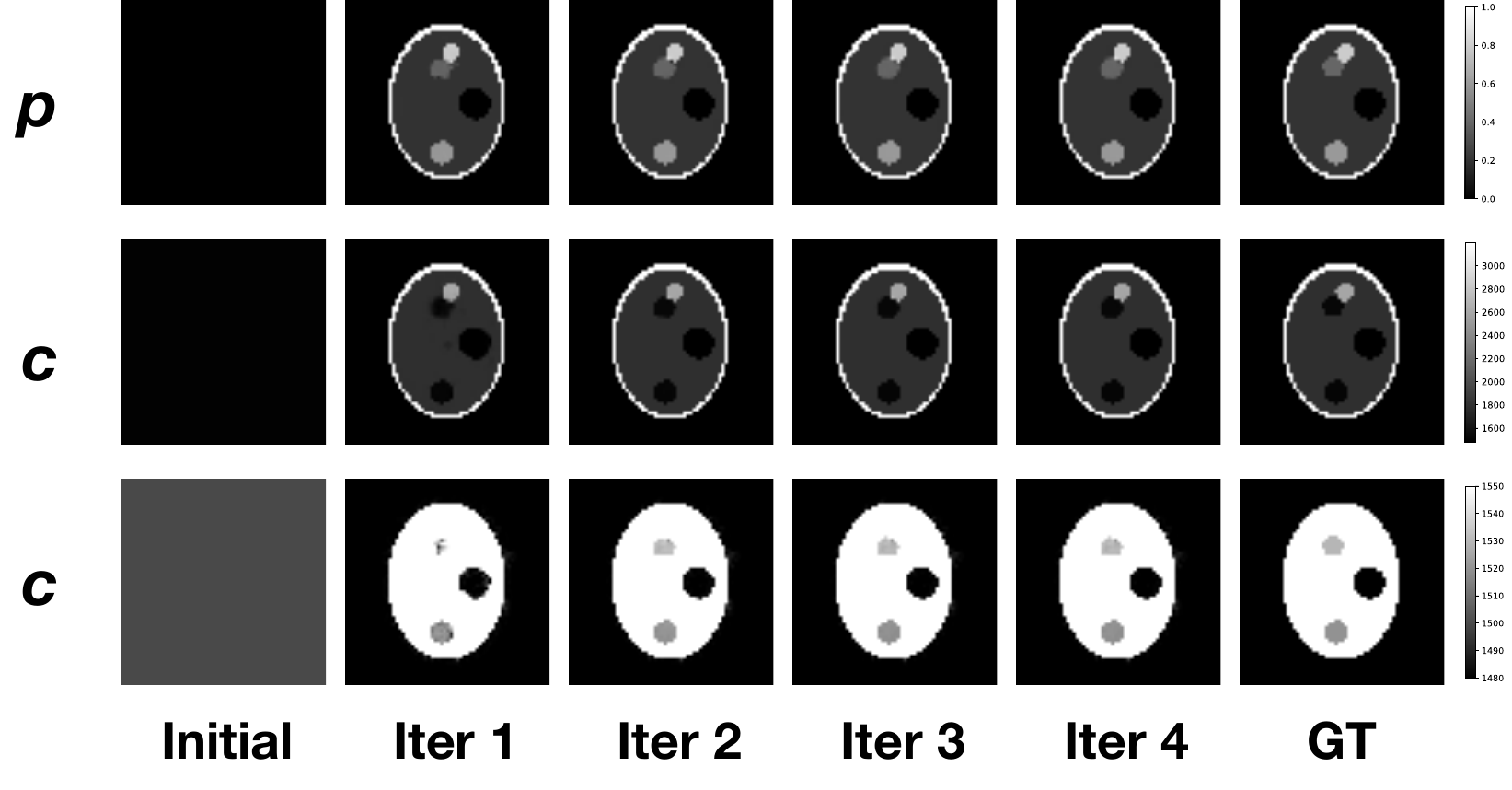}
    \caption{The first case of the simultaneously reconstructed results by applying SR-Net to the same phantom. The first row shows the initial pressure. The second and third rows show the sound speed distribution in range of [1480, 3198] and of [1480, 1550], respectively. GT means the ground-truth.}
    \label{fig:case_508}
\end{figure}

\subsection{Experimental results}

Throughout the experiments, the initial pressure was initialized uniformly as zero and the sound speed was initialized uniformly as 1,500 m/s. The maximum number of iteration $k_{max}$ was set to 4. The gradient of the data fidelity with respect to the initial pressure is computed by the adjoint state method using the $k$-wave package~\cite{Treeby2010}. The proposed SR-Net was implemented by Pytorch and trained on four NVIDIA 1080Ti GPUs.

\begin{figure}[htbp]
    \centering
    \includegraphics[width=1.0\textwidth]{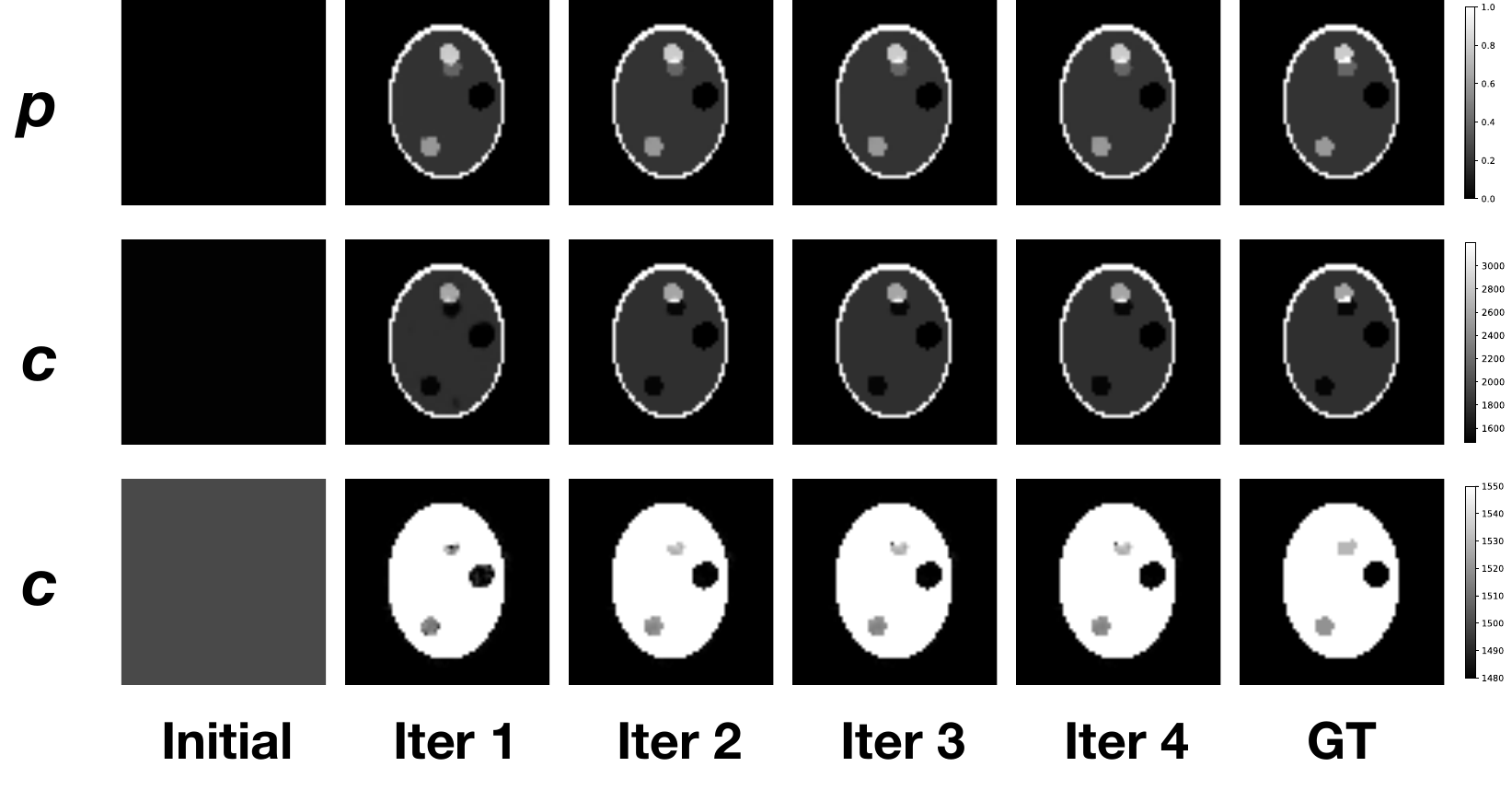}
    \caption{The second case of the simultaneously reconstructed results by applying the SR-Net to the same phantom. The first row shows the initial pressure. The second and third rows show the sound speed distribution in the range of [1480, 3198] and of [1480, 1550], respectively. GT means the ground-truth.}
    \label{fig:case_90}
\end{figure}

We show two testing images in Figs.~\ref{fig:case_508} and \ref{fig:case_90}, which were not used in training stage. It is observed that the proposed method can reconstruct the initial pressure and sound speed simultaneously with very high accuracy. Mean absolute error for sound speed and initial pressure were calculated for each reconstructed image at each iteration. The average value and standard deviations of these errors are contained in Table~\ref{table_mae}.

\begin{table}[htbp]
\centering
\caption{Mean absolute error of the initial pressure and sound speed distribution for each iteration in the form of $\mathrm{MEAN} \pm \mathrm{STD}$.}\label{table_mae}
 \begin{tabular}{c c c c c} 
 \hline
\toprule
 Mean Absolute Error & Iter 1 & Iter 2 & Iter 3 & Iter 4 \\ 
 \midrule
 $p$ & (1.43$\pm$0.36)$\times 10^{-2}$ & (0.86$\pm$0.33)$\times 10^{-2}$ & (0.85$\pm$0.33)$\times 10^{-2}$ & (0.85$\pm$0.33)$\times 10^{-2}$\\ 
 $c$ [m/s] & 3.13$\pm$0.64 & 1.33$\pm$0.52 & 1.25$\pm$0.52 & 1.24$\pm$0.52 \\
 \bottomrule
\end{tabular}
\end{table}

\subsection{Generalization ability of the trained SR-Net on a new phantom}

\begin{figure}[htbp]
    \centering
    \includegraphics[width=1.0\textwidth]{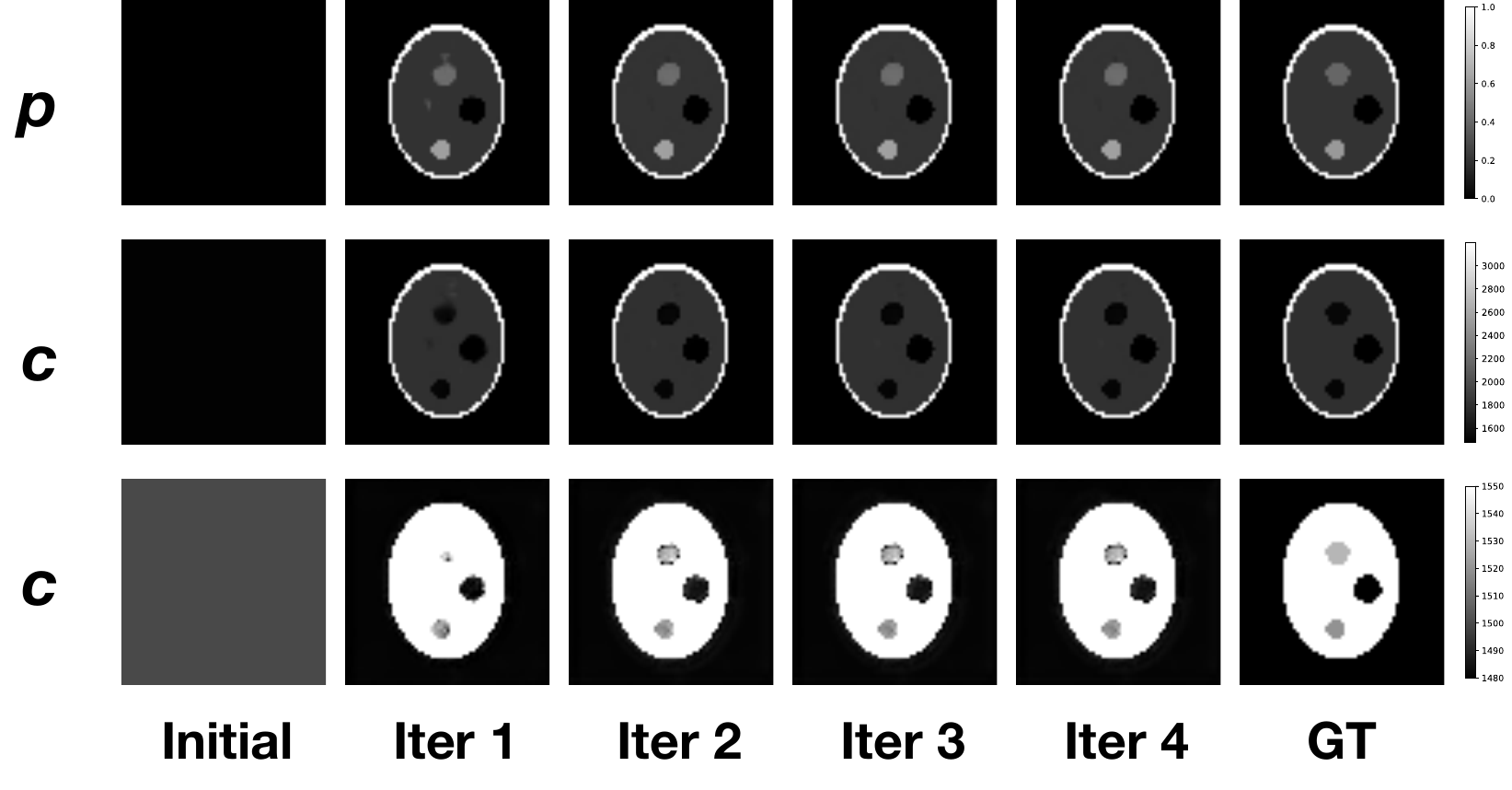}
    \caption{The first case of the simultaneously reconstructed results by applying the SR-Net to the new phantom. The first row shows the initial pressure. The second and third rows show the sound speed distribution in the range of [1480, 3198] and of [1480, 1550], respectively. GT means the ground-truth.}
    \label{fig:case_2}
\end{figure}

\begin{figure}[htbp]
    \centering
    \includegraphics[width=1.0\textwidth]{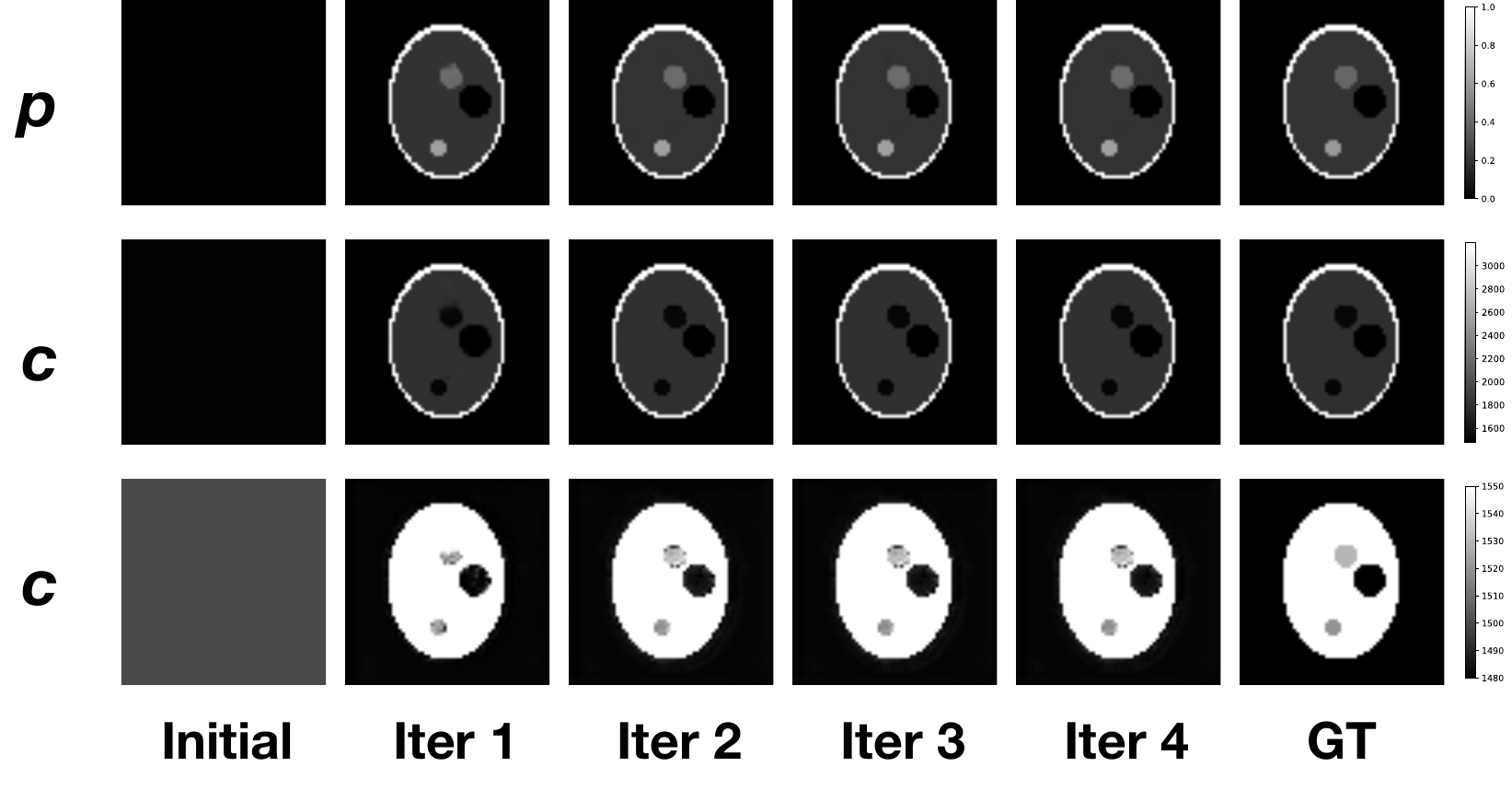}
    \caption{The second case of the simultaneously reconstructed results by applying the SR-Net to the new phantom. The first row shows the initial pressure. The second and third rows show the sound speed distribution in the range of [1480, 3198] and of [1480, 1550], respectively. GT means the ground-truth.}
    \label{fig:case_3}
\end{figure}

All Shepp-Logan phantoms used to train the model featured all six regions listed in Table~\ref{tab:index}. To test the generalizability of the model, we reconstructed four phantoms with region 5 removed. Figures~\ref{fig:case_2} and~\ref{fig:case_3} illustrate two cases of this test. From these cases, it can be observed that the model still accurately reconstructs phantoms that contain less regions than the phantoms used to train the model.

\section{Discussions and Conclusion}

The main weakness of this study is the numerical nature, which may not fully reflect all the physical factors in real clinical/pre-clinical applications. We plan to perform physical phantom experiments as a next step.  Also, we are interested in animal studies in vivo. Nevertheless, the numerical results and the network trained with simulated data may serve a baseline for further improvements. Hopefully, this imaging modality may find practical use in some important tasks such as breast imaging.

In conclusion, we have developed a simultaneous reconstruction network (SR-Net) to jointly recover the sound speed and initial pressure in photoacoustic tomography. The proposed SR-Net has several merits:  automatically learning the step size, regularization term, and the trade-off between data fidelity and regularization in a data-driven manner, and not requiring the gradient of the data fidelity with respect to sound speed. In the future, we will further improve the structure of the network and also validate it on more complicated datasets.

\section{Acknowledgment}

The research of Y. Yang is partly supported by the NSF grant DMS-1715178, the AMS-Simons travel grant, and the startup fund from Michigan State University. The authors would like to thank NVIDIA Corporation for the
donation of GPU used for this research.

\end{document}